\documentclass[iop, apjl]{emulateapj}
\usepackage{graphicx}
\usepackage{epsf}
\usepackage{epsfig}
\usepackage{amssymb,amsmath}
\usepackage[usenames]{color}
\usepackage{times}
\usepackage[colorlinks=true,linktocpage=false,citecolor=black,linkcolor=blue,urlcolor=blue]{hyperref}
\usepackage[nolist,nohyperlinks]{acronym}
\usepackage{xspace}
\usepackage{breakurl}
\usepackage{natbib}
\def\be{\begin{equation}}
\def\ee{\end{equation}}
\def\beq{\begin{eqnarray}}
\def\eeq{\end{eqnarray}}
\def\bi{\begin{itemize}}
\def\ei{\end{itemize}}
\def\ben{\begin{enumerate}}
\def\een{\end{enumerate}}
\def\nsbh{\ac{NS}-\ac{BH}\xspace}
\def\bns{\ac{NS}-\ac{NS}\xspace}
\newcommand{\mBH}{M_{\rm BH}}
\newcommand{\mNS}{M_{\rm NS}}
\newcommand{\spin}{\chi_{\rm BH}}
\newcommand{\zspin}{\chi_{\hat L,{\rm BH}}}
\newcommand{\mChirp}{M_{\rm Chirp}}
\newcommand{\mDisk}{M_{\rm disk}}
\newcommand{\mThresh}{M_{\rm Threshold}}

\shorttitle{NS-BH GW and EM Prospects}

\begin{document}

\title{Prospects for joint gravitational-wave and electromagnetic
  observations of neutron-star--black-hole coalescing binaries}

\author{Francesco Pannarale}%
\author{Frank Ohme}%
\affil{%
  School of Physics and Astronomy, Cardiff University, The Parade,
  Cardiff CF24 3AA, UK; \href{mailto:francesco.pannarale@ligo.org}{francesco.pannarale@ligo.org}, \href{mailto:frank.ohme@ligo.org}{frank.ohme@ligo.org}
}%

\begin{abstract}
  Coalescing neutron-star--black-hole (\acused{NS}\acused{BH}\nsbh)
  binaries are a promising source of \ac{GW} signals detectable with
  large-scale laser interferometers such as Advanced \acl{LIGO} and
  Virgo. They are also one of the main \ac{SGRB} progenitor
  candidates. If the \ac{BH} tidally disrupts its companion, an
  \ac{SGRB} may be ignited when a sufficiently massive accretion disk
  forms around the remnant \ac{BH}. Detecting an \nsbh~coalescence
  both in the \ac{GW} and \ac{EM} spectrum offers a wealth of
  information about the nature of the source. How much can actually be
  inferred from a joint detection is unclear, however, as a mass/spin
  degeneracy may reduce the \ac{GW} measurement accuracy. To shed
  light on this problem and on the potential of joint \ac{EM}+\ac{GW}
  observations, we here combine recent semi-analytical predictions for
  the remnant disk mass with estimates of the parameter-space portion
  that is selected by a \ac{GW} detection. We identify cases in which
  an \ac{SGRB} ignition is supported, others in which it can be
  excluded, and finally others in which the outcome depends on the
  chosen model for the currently unknown \ac{NS} equation of state. We
  pinpoint a range of systems that would allow us to place lower
  bounds on the equation of state stiffness if both the \ac{GW}
  emission and its \ac{EM} counterpart are observed. The methods we
  develop can broaden the scope of existing \ac{GW} detection and
  parameter-estimation algorithms and could allow us to disregard
  about half of the templates in an \nsbh~search following an
  \ac{SGRB} trigger, increasing its speed and sensitivity.
\end{abstract}



\maketitle

\begin{acronym}
\acrodef{1D}[1D]{one-dimansional}
\acrodef{BH}[BH]{black hole}
\acrodef{EM}[EM]{electromagnetic}
\acrodef{EOS}[EOS]{equation of state}
\acrodef{GW}[GW]{gravitational-wave}
\acrodef{KAGRA}[KAGRA]{Kamioka Gravitational wave detector}
\acrodef{LIGO}[LIGO]{Laser Interferometer Gravitational-Wave Observatory}
\acrodef{NS}[NS]{neutron star}
\acrodef{PCA}[PCA]{principal component analysis}
\acrodef{PN}[PN]{post-Newtonian}
\acrodef{SGRB}[SGRB]{short gamma-ray burst}
\acrodef{ShortGRB}[SGRB]{Short gamma-ray burst}
\acrodef{SNR}[SNR]{signal-to-noise ratio}
\end{acronym}

\section{Introduction}
\acresetall
\acp{ShortGRB}\acused{SGRB} are brief, intense, non-repeating flashes
of radiation that, while active, outshine every other source in the
gamma-ray sky. During the last decade, afterglow observations,
host-galaxy identifications, and redshift measurements greatly
augmented our understanding of \acp{SGRB}, but the exact nature of
their progenitors is still unknown \citep{Berger2011}. The current
consensus is that the observed properties of (a subset of) \acp{SGRB}
may be ascribed to matter accretion onto a stellar mass \ac{BH} or
onto a compact object that evolves into one.

The coalescence of compact binaries comprising \acp{NS} --- \nsbh~and
\bns~systems --- is therefore a natural astrophysical scenario for the
production of viable \ac{SGRB} progenitor candidates
\citep{Nakar:2007yr}, as the coalescence remnant may consist of a
\ac{BH} with negligible baryon contamination along its rotation axis,
surrounded by a hot, massive accretion disk (e.g.,
\citet{Foucart2013a,Rezzolla:2011}) that releases energy while
accreting onto the \ac{BH}.

With the imminent start of the era of second-generation
laser-interferometric \ac{GW} detectors, further light may be shed on
compact binaries as \ac{SGRB} progenitors. The upgraded detectors of
the \acl{LIGO} (LIGO\acused{LIGO}; \citet{AdvLIGO}) and Virgo
\citep{AdvVirgo} collaborations, and the newly built \ac{KAGRA}
\citep{KAGRA} will allow for directly observing the last instants of
the evolution of \ac{NS} systems via the detection of their \ac{GW}
emission, providing unprecedented information about the physical
nature of these systems.

In this Letter, we focus on \nsbh~systems, noting that a \ac{GW}
observation of these binaries would constitute their first direct
detection, as opposed to already observable \bns~systems
\citep{Kiziltan2013}. At its design sensitivity ($\sim2019$;
\citet{LSC2013}), Advanced LIGO is expected to detect $0.03$--$4.4$
\nsbh~mergers per year in single detector mode, while the expected
detection rate for a network of three detectors operating at this
sensitivity is $0.07$--$9.4$ events per year \citep{DominikPaper3}.

Within the \ac{GW} community, several \nsbh~binary studies focused on
the signatures left by the currently unknown \ac{NS} \acl{EOS}
(EOS\acused{EOS}; \citet{Lattimer07}) on the emitted \ac{GW} signal
and on the possibility of constraining the \ac{EOS} (e.g.,
\citet{Kyutoku2010, Pannarale2011, Pannarale2013a, Lackey2012,
  Lackey2013, Wade2014}). Our analysis considers only the inspiral
portion of the \ac{GW} signal, since a point-particle description of
this signal is accurate enough for detection purposes, as shown by
\citet{Pannarale2011}. However, we go beyond the \ac{GW} detection
scenario and examine the case of \emph{joint} (inspiral)
\ac{GW}-\ac{EM} detections. We show where in the \nsbh~parameter space
such detections are possible and address the idea of using them to
constrain the \ac{NS} \ac{EOS}. Additionally, we discuss that the
framework we develop would allow us to (1) assess the importance of an
\ac{EM} follow-up to a \ac{GW} detection, and (2) to improve the
performance of offline \ac{GW} searches following \ac{SGRB} triggers.

Recently, \citet{Maselli2014} addressed similar issues, but with a
different scope. Assuming \ac{GW} measurements of the parameters of
non-precessing binaries, including the \ac{NS} tidal deformability,
they estimated the accuracy with which these parameters are determined
by a linear order Fisher-matrix approximation. Consistently with this
approximation, they used a Gaussian probability distribution to infer
the chances of finding a coincident \ac{EM} merger signature. We go
beyond this study by identifying parameter degeneracies in the \ac{GW}
measurement through a \ac{PCA}: this enables us to present a
\emph{complete} analysis of the relevant \nsbh parameter space without
relying on a few cases and high \acp{SNR}. Additionally, we ensure
that all our estimates are conservative, particularly when addressing
precessing \ac{BH} spins, so that we identify parameter-space regions
where the mere existence of a coincident \ac{GW}-\ac{EM} detection
allows us to put lower bounds on the \ac{NS} \ac{EOS} stiffness,
without directly measuring any tidal deformability and even in the
presence of large uncertainties. Conversely, we identify regions where
the merger dynamics does not support any \ac{EM} emission. Relating
this parameter-space portion to the regions covered by aligned-spin
templates, we suggest removing templates with no potential
\ac{EM}-counterpart to optimize \ac{SGRB}-targeted \ac{GW} searches.

\section{Methodology}
\subsection{Gravitational-wave Measurement} \label{sec:GW_measurement}
The most sensitive method to detect \acp{GW} of known signature in
noise-dominated instrument data is to cross correlate the data with
theoretically predicted waveforms. This relies on accurate
descriptions of the expected signals. We model compact binary inspiral
waveforms by a standard frequency-domain \ac{PN} approximant
\citep{Damour:2000zb} in the form used by \citet{Ohme2013}. Even if
the model were to perfectly describe the expected signal, which we
assume here for simplicity, the accuracy of inferred binary parameters
is limited by the fact that differences between signals from distinct
sources may be undetectable below the inevitable noise floor. Thus,
\ac{GW} measurements only select a \emph{parameter-space region}
consistent with an observation; we shall summarize the method we
employ to identify this region below. Note that we refer to the
fiducial signal as ``target signal,'' whereas ``templates'' denote
waveform models used to analyze the data.

Various approaches estimate the accuracy of \ac{GW} measurements,
ranging from linear-order approximations (see
\citet{Vallisneri:2007ev} for details on the Fisher-information-matrix
approach) to multi-dimensional numerical analysis techniques (see
\citet{Aasi:2013jjl} for recent results). In this Letter we identify
parameter degeneracies through a semi-analytical \ac{PCA}, as outlined
in \citet{Ohme2013}. This method was shown to be more accurate than
simple linear-order approximations, and was successfully implemented
in \ac{GW} search algorithms as an integral part of the template bank
construction \citep{Harry:2013tca}. While only the above mentioned
numerical techniques take advantage of the full detector data
information, our \ac{PCA} is sufficiently accurate to demonstrate the
general idea presented in this Letter and simultaneously allows us to
analyze a large parameter space.

The details of the algorithm are as follows. We consider an
\nsbh~\ac{GW} source and parameterize it by the constituent masses
$\mBH$ and $\mNS$ and the \ac{BH} spin. We neglect the \ac{NS} spin as
it is expected to be small in compact binaries
\citep{Bildsten:1992my,Kochanek:1992wk}. For templates, we also
exclude precession by taking the \ac{BH} spin to be aligned with the
total orbital angular momentum, $\hat{L}$, with a dimensionless spin
projection $\zspin\in[-1,1]$. This restriction will be relaxed for
target signals as discussed in the following section. As all other
source parameters, such as distance, sky location, polarization,
hardly correlate with the parameters we consider, we leave them to
dedicated studies (e.g., \citet{Singer:2014qca}).

We then perform a \ac{PCA} with the frequency-domain \ac{PN} waveform
model, as implemented by \citet{Ohme2013}, assuming the Advanced LIGO
zero-detuned, high-power mode design sensitivity \citep{LIGO-noise},
with $15\,{\rm Hz}$ lower cutoff. The high-frequency cutoff is given
by the frequency of the last stable circular orbit around a
Schwarzschild \ac{BH} with mass $\mBH+\mNS$. The \ac{PCA} provides the
parameter-space directions that are well constrained by \ac{GW}
observations. We take the first two of these principal components and
assume them to be measured exactly at the values of our fiducial
target system, which reduces the three-dimensional parameter space
$(\mNS,\mBH,\zspin)$ to a \ac{1D} line consistent with the target.

The result we obtain is interpreted as follows. Given a \ac{GW}
detection, we can measure two values very accurately. To a good
approximation, the first one is described by the \emph{chirp mass:}
\begin{equation}
  \mChirp=\frac{(\mNS\;\mBH)^{3/5}}{(\mNS+\mBH)^{1/5}}, 
\end{equation}
whereas the second one restricts the range of consistent parameters to
a narrow line in the mass-ratio/spin space (at constant $\mChirp$).

We neglect \ac{SNR}-dependent uncertainties on these two best-measured
parameter combinations, as they are too small to affect our analysis.
Specifically, all \ac{GW}-degenerate lines we shall show below are
clearly distinguishable at any \ac{SNR} sufficient for detection (see
Section III in \citet{Ohme2013} for details). Incorporating third
principal component information would restrict the \ac{1D} line to a
finite length, but we make the conservative assumption that this would
not be more restrictive than the physical bounds imposed by cosmic
censorship ($\vert\spin\vert\leq1$) and the plausible \ac{NS} mass
range. As a technical caveat, note that although the functional form
of the principal components only weakly depends on the parameter-space
point one is considering, a total-mass dependence is inherited from
the upper cutoff frequency: we take care of this by re-calculating the
principal components for a range of target sources.

\subsection{Electromagnetic Counterpart}

We use the baryonic mass $\mDisk$ of the disk remnant of
\nsbh~binaries as a proxy for \ac{EM} counterparts: if $\mDisk$
exceeds a threshold mass $\mThresh$, then the merger can produce a
counterpart, albeit not necessarily a detectable one, otherwise it is
treated as electromagnetically silent. We determine $\mDisk$ through
the fit of \citet{Foucart2012} to numerical-relativity disk-mass
predictions for aligned \nsbh~mergers (Equations (7), (8), (12), and
(13) therein). For misaligned mergers, we generalize this formula as
suggested by \citet{Stone2013} (see text supporting Equation (9)
therein). While our approach relies on this fit, any alternative and
possibly improved method could be used in future applications. Unless
otherwise noted, we set $\mThresh=0.03M_\odot$: this is on the low end
of the disk-mass values expected to allow for \acp{SGRB} $1$ s in
duration (e.g., \citet{Stone2013}, end of Section III).

\begin{figure*}
  \begin{center}
    \includegraphics[width=\columnwidth,clip=true]{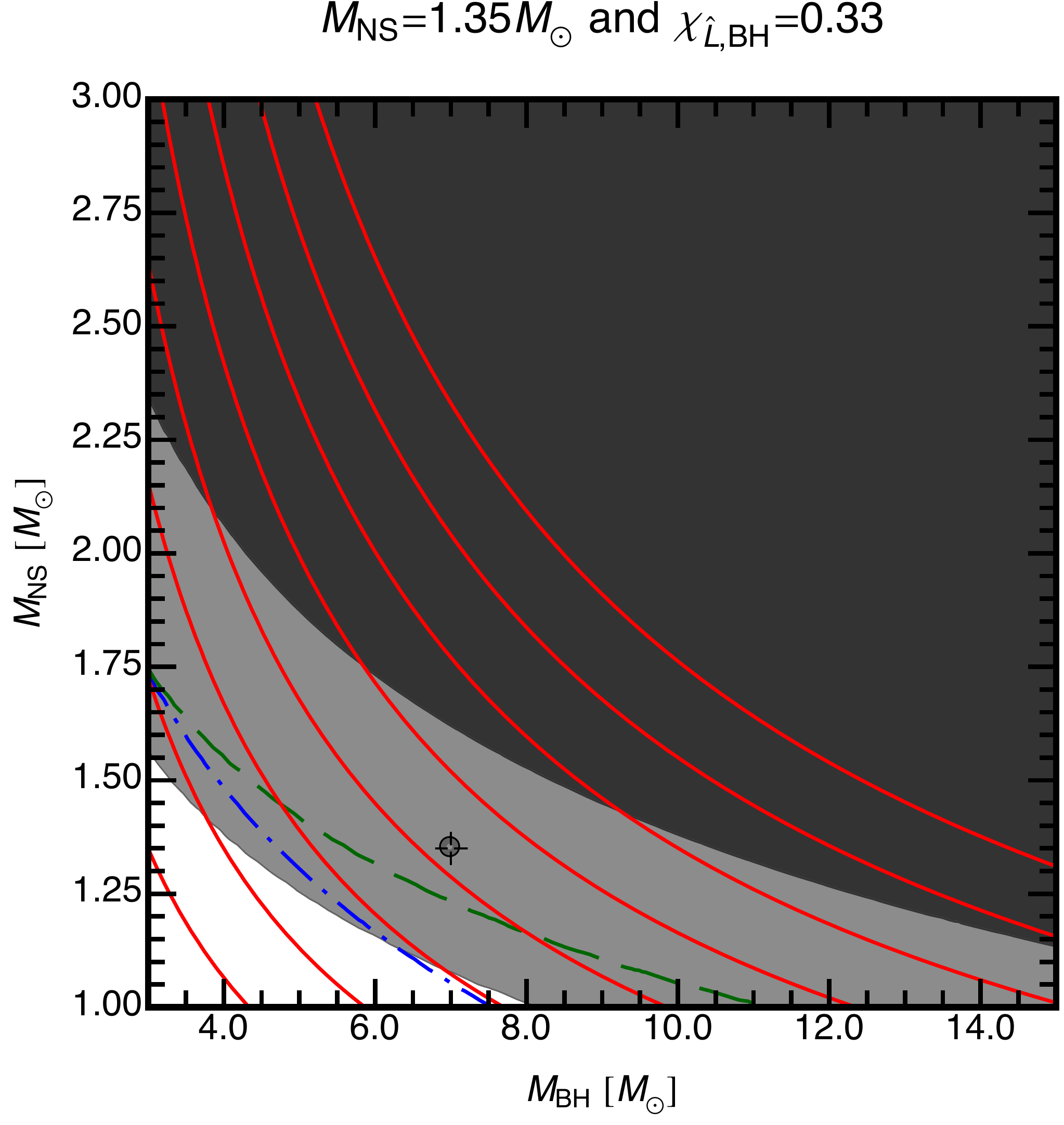}
    \hskip 1em
    \includegraphics[width=\columnwidth,clip=true]{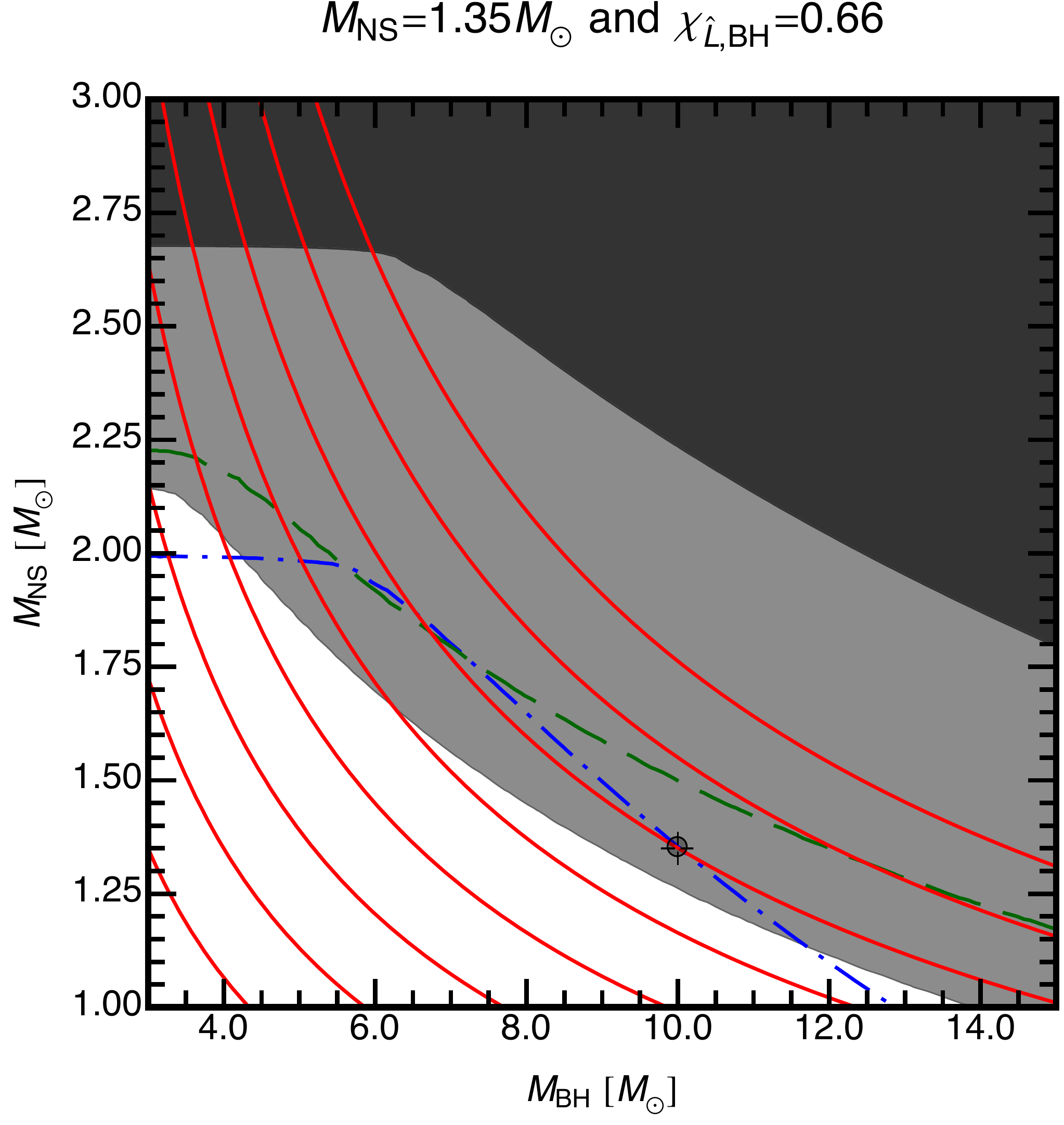}
    \caption{\ac{GW} degeneracies, illustrated by the first principal
      component lines (red) in the $\mBH$--$\mNS$ plane, for target
      systems with parameters defined in the title of each panel; in
      the dark gray area no \ac{EM} counterpart is possible, in the
      light gray area the presence of an \ac{EM} counterpart depends
      on the \ac{NS} \ac{EOS}, while in the white region the
      counterpart is guaranteed by all \ac{NS} \acp{EOS}. Results for
      APR2 and for strange quark matter (SQM3) are shown in green
      dashed and blue dot-dashed lines,
      respectively. \label{fig:realisticEOSs}}
  \end{center}
\end{figure*}

The following procedure combines \ac{EM} prospects with \ac{GW}
measurements. For each fiducial target signal, we calculate $\mDisk$
for all sources selected by a \ac{GW} measurement (see
Section \ref{sec:GW_measurement}). In addition to the parameters
$\mNS$, $\mBH$, and $\zspin$, we consider various \ac{NS} \acp{EOS}
and allow the \ac{BH} spin of the target signal to be tilted with
respect to $\hat{L}$.  We make the conservative assumption that
precessing systems, if detected, are identified by the non-precessing
template that correctly captures the secular inspiral rate. Following
\citet{Schmidt:2012rh} and \citet{Pekowsky:2013ska}, we define this
template by the same mass parameters as the signal and the projection
of the generic spin onto $\hat{L}$.

We will quote results for the WFF1 \citep{Wiringa88}, PS
\citep{PandharipandeSmith:1975}, APR2 \citep{Akmal1997,Akmal1998a},
and SQM3 \citep{Witten1984,Farhi1984,Alcock1988} \acp{EOS}. Although
superseded, the first two \acp{EOS} yield extremely compact and
extremely large NSs, respectively, generating results that bracket the
ones produced by all other \acp{EOS}. APR2, on the other hand, is the
most complete nuclear many-body \ac{EOS} to date and is supported by
current nuclear physics and astrophysical constraints: it may be
thought of as a reference for our \ac{NS} \ac{EOS} best
guess. Finally, SQM3 is an exotic \acp{EOS} employed to illustrate the
case of strange quark stars; however, no strange-star--\ac{BH} merger
numerical simulations are available, so this is untested territory for
the $\mDisk$ fit.

Given an \ac{EOS}, the techniques just described allow us to (1) split
the parameter space into an \ac{EM}-silent region (where
$\mDisk<\mThresh$) and an \ac{EM}-loud one (where $\mDisk>\mThresh$),
and to (2) study \ac{GW} degeneracies throughout the parameter space.

\section{Results}
Given a set of \nsbh~parameters, as the \ac{NS} \ac{EOS} is softened
and the \ac{NS} radius decreases, $\mDisk$ decreases due to the
increased difficulty in tidally disrupting the \ac{NS}. The volume of
the (potentially) \ac{EM}-loud parameter-space region decreases
consequently. Similarly, the stiffer the \ac{EOS}, the higher the
chances of having an \ac{EM} counterpart \citep{Pannarale2010}.

A second, less intuitive aspect is related to the \ac{BH} spin.  All
``\ac{GW} parameters'' $(\mNS,\mBH,\zspin)$ being fixed, a higher
\ac{BH} spin magnitude yields a higher $\mDisk$, because increasing
$\spin$ shrinks the innermost stable orbit more rapidly than the tidal
disruption orbit: the greater the difference between the radii of
these two orbits, the higher $\mDisk$. We remark that this statement
on the dependency of $\mDisk$ on $\spin$ is more general than ones for
aligned mergers, as we allow for the \ac{BH} spin to be tilted with
respect to $\hat{L}$. While we assume that systems with equal values
of $\mNS$, $\mBH$, and $\zspin$ emit similar \ac{GW} signals, the
chances of having \ac{EM} counterparts depend on $\spin$: setting
$\spin=0.998$ \citep{Thorne1974} in our analysis therefore allows us
to make \emph{conservative} statements about the \ac{EM}
counterpart. In other words, if $\mDisk<\mThresh$ for $\spin=0.998$
and a given set of $\mNS$, $\mBH$, and $\zspin$ values, no \ac{EM}
emission is expected to be associated with \nsbh~mergers with the same
mass and aligned-spin component values. Bearing this in mind, from
here onward we will consider $\spin=0.998$ only.

\begin{figure*}
  \begin{center}
    \includegraphics[width=\columnwidth,clip=true]{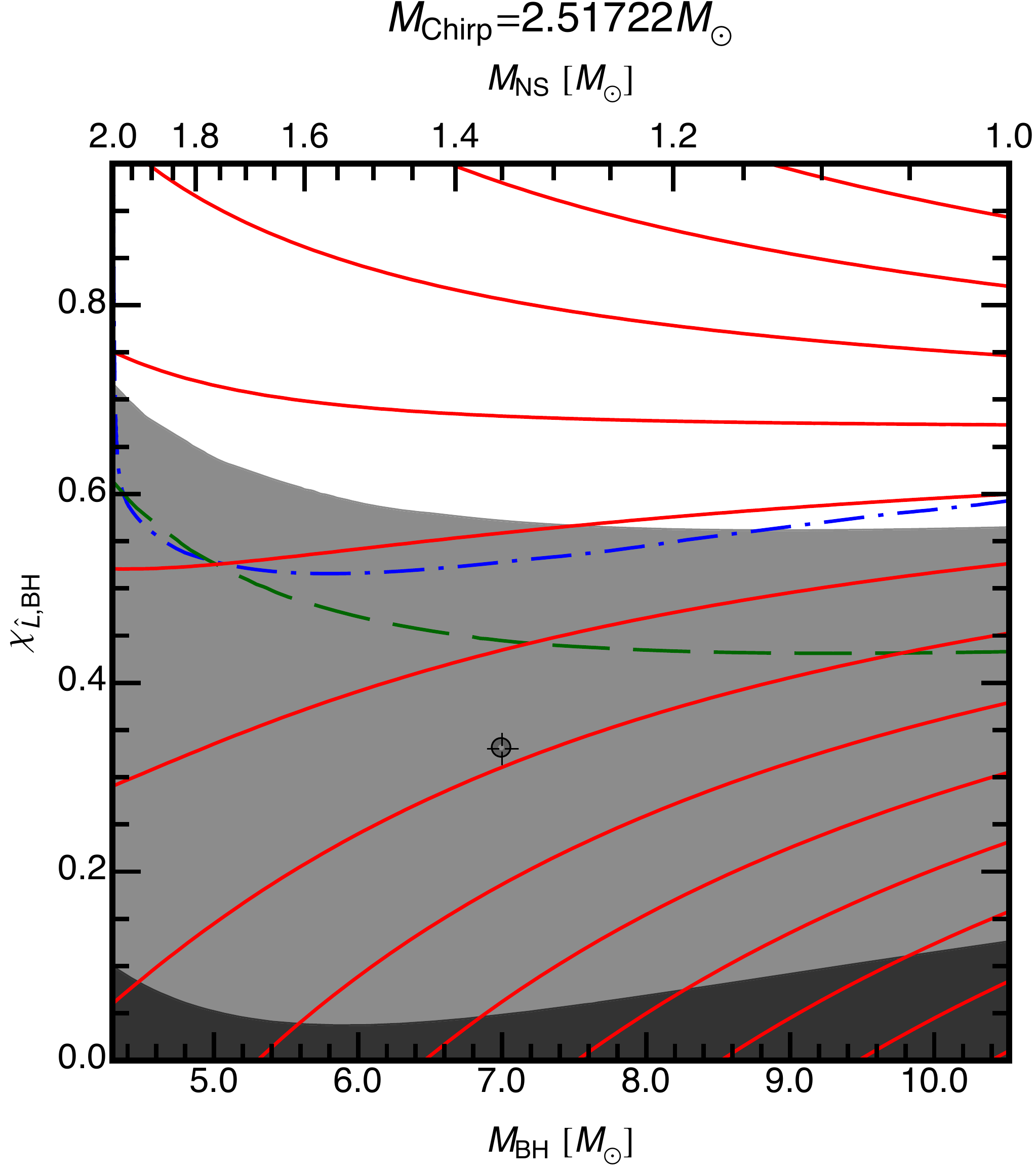}
    \hskip 1em
    \includegraphics[width=\columnwidth,clip=true]{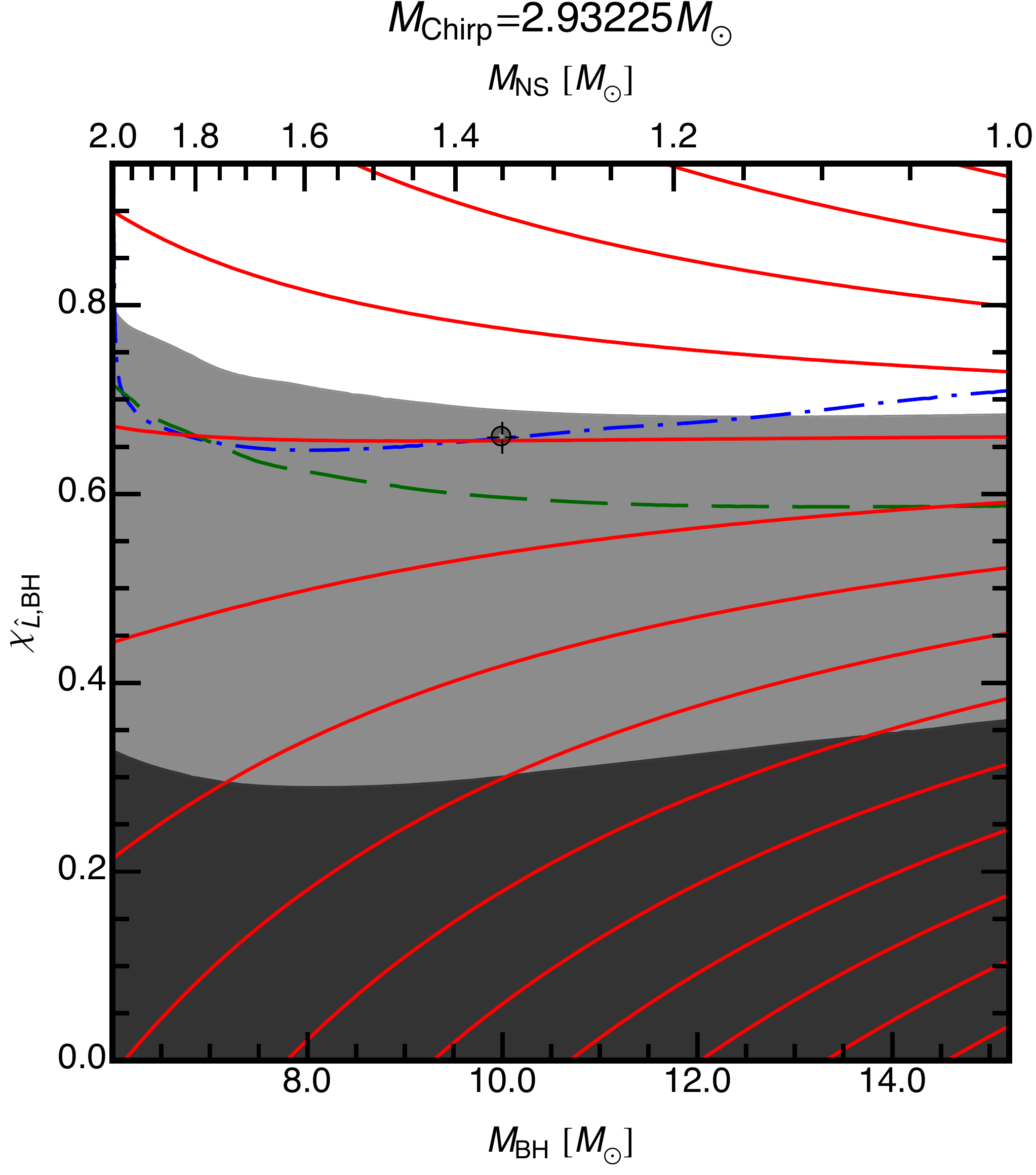}
    \caption{Second principal component lines in the $\mBH$--$\zspin$
      plane for constant $\mChirp$, as defined in the title of each
      panel (see Figure \ref{fig:realisticEOSs} for the color
      code). \label{fig:realisticEOSs2}}
  \end{center}
\end{figure*}

Figures \ref{fig:realisticEOSs} and \ref{fig:realisticEOSs2} are
representative of our results. In Figure \ref{fig:realisticEOSs}, we
pick a family of target systems with $\mNS=1.35M_\odot$ and
$\zspin=0.33$($0.66$) in the left(right) panel. We then span the
template $\mBH$--$\mNS$ plane and illustrate the regions selected by
\ac{GW} measurements as red solid curves. These are lines of constant
first principal component, essentially constant $\mChirp$, along which
the template $\zspin$ varies. Along a given curve, a lower (higher)
template $\mBH$ requires a higher (lower) $\mNS$ to preserve
$\mChirp$, and a lower(higher) $\zspin$ to ensure a high match between
the template and the target that lives on that same curve.

The dark area in the background indicates that no \ac{EM} counterpart
is available, because $\mDisk<\mThresh$ for any \ac{EOS}. On the
boundary between the two gray areas, $\mDisk=\mThresh$ for the PS
\ac{EOS}: any \nsbh~binary below it is expected to have an \ac{EM}
counterpart if we assume the PS \ac{EOS} to be valid. Notice that the
maximum \ac{NS} mass can also contribute to shaping this curve: this
is the case of the flattening at $\mNS\simeq2.66M_\odot$ in the right
panel.

On the boundary between the light gray and the white areas
$\mDisk=\mThresh$ for the WFF1 \ac{EOS}. Therefore, in the light gray
region, $\mDisk$ can exceed $\mThresh$, depending on the \ac{NS}
\ac{EOS}. Within this region, we explicitly show the $\mDisk=\mThresh$
curve for the ``standard'' APR2 \ac{EOS} in dashed green.

The white region denotes \nsbh~binaries with $\mDisk>\mThresh$ for any
\ac{EOS}. Finally, the dot-dashed blue line marks the
$\mDisk=\mThresh$ boundary for strange quark stars (SQM3
\ac{EOS}). Once more, the maximum mass supported by the \ac{EOS}
flattens the curve for moderate-to-high $\zspin$ values.

Figure \ref{fig:realisticEOSs2} uses the same color code. Here we
consider a fixed $\mChirp$ target-system family and show the behavior
of the second \ac{GW} principal component in the template
$\mBH$--$\zspin$ plane. The $\mChirp$ value corresponds to a
$1.35M_\odot+7(10)M_\odot$ system in the left(right) panel.  On a
given side of Figures \ref{fig:realisticEOSs} and
\ref{fig:realisticEOSs2}, crosshairs denote the one target binary with
$\mNS$, $\zspin$, and $\mChirp$ as specified by the two panel titles
on that side. This allows us to visually follow a distinct binary when
varying $\zspin$ and $\mChirp$.

A number of observations can be made. There are parameter-space
regions in which joint \ac{GW}-\ac{EM} detections are clearly possible
(white), others in which they are not possible (dark gray), and
finally ones in which this possibility is \ac{EOS}
dependent. Quantitative information of this kind would be valuable
when deciding whether to send an \ac{EM} alert following an \nsbh
\ac{GW} detection, or when performing a \ac{GW}~follow-up to an
\ac{SGRB} trigger (see next section).

We find that increasing(decreasing) the target $\zspin$($\mChirp$)
enhances the chances of having an \ac{EM} counterpart, as a larger
portion of the parameter-space slice is \ac{EM}-active. This
generalizes the well-established behavior of $\mDisk$ (i.e., it grows
with the \ac{BH} spin magnitude in aligned-spin binaries, but
decreases for increasing $\mBH$) and translates it into terms relevant
to \ac{GW} science.
  
Interestingly, just a few principal component lines cut through the
entire \ac{EOS}-sensitive region. Joint \ac{GW}-\ac{EM} detections can
thus constrain the \ac{NS} \ac{EOS}, despite the \ac{GW} measurement
degeneracies. More specifically, a joint detection could potentially
place a lower bound on the \ac{EOS} stiffness. If an \nsbh~\ac{GW}
signal is detected, the source $\mChirp$ and second principal
component can be determined. The presence of, say, an \ac{SGRB}
counterpart would indicate that part of the \ac{NS} matter survived
the merger, thus reducing the size of the light gray area in our
Figure \ref{fig:realisticEOSs} and \ref{fig:realisticEOSs2}
examples. This effectively places a lower limit on the \ac{NS}
compactness and hence on the \ac{EOS} stiffness.
  
A direct consequence of the previous point is that a single joint
detection could exclude the presence of a strange star in the
progenitor. As an example, if the principal component values of the
target we fixed in the left panels of Figures \ref{fig:realisticEOSs}
and \ref{fig:realisticEOSs2} are measured, and an \ac{SGRB} signal is
detected, this could not be supported by the SQM3 \ac{EOS}.

\section{Discussion}
In this Letter we considered the \ac{GW} and \ac{EM} emission of
\nsbh~binaries. We used \ac{PCA} techniques for the \ac{GW} inspiral
signal and a proxy to determine the presence/absence of an \ac{EM}
counterpart. We explained how joint \nsbh~binary merger detections can
place lower limits on the \ac{NS} \ac{EOS} stiffness that constrain
the \ac{EOS}, potentially excluding strange quark matter \acp{EOS}
with a single joint detection. Since our statements are mainly based
on semi-analytical expressions derived from either numerical
simulations or a \ac{PCA}, this analysis can be performed quickly,
making it an ideal addition to existing \ac{GW} search and
parameter-estimation pipelines.

Our method may also be used as a framework to assess the importance of
sending alerts for \ac{EM} follow-ups to \nsbh~\ac{GW}
detections. Conversely, a speed-up of offline \nsbh~\ac{GW} searches
in coincidence with \ac{SGRB} triggers is possible: the number of
templates used in the search may be reduced by identifying
\nsbh~systems that no \ac{EOS} supports as \ac{SGRB} progenitors. This
would also improve the search sensitivity, as keeping templates that
cannot be associated with an \ac{SGRB} may harm the search by
increasing the false-alarm rate, which only higher \ac{SNR} thresholds
could counteract.

\begin{figure}
  \centering
  \includegraphics[width=\columnwidth,clip=true]{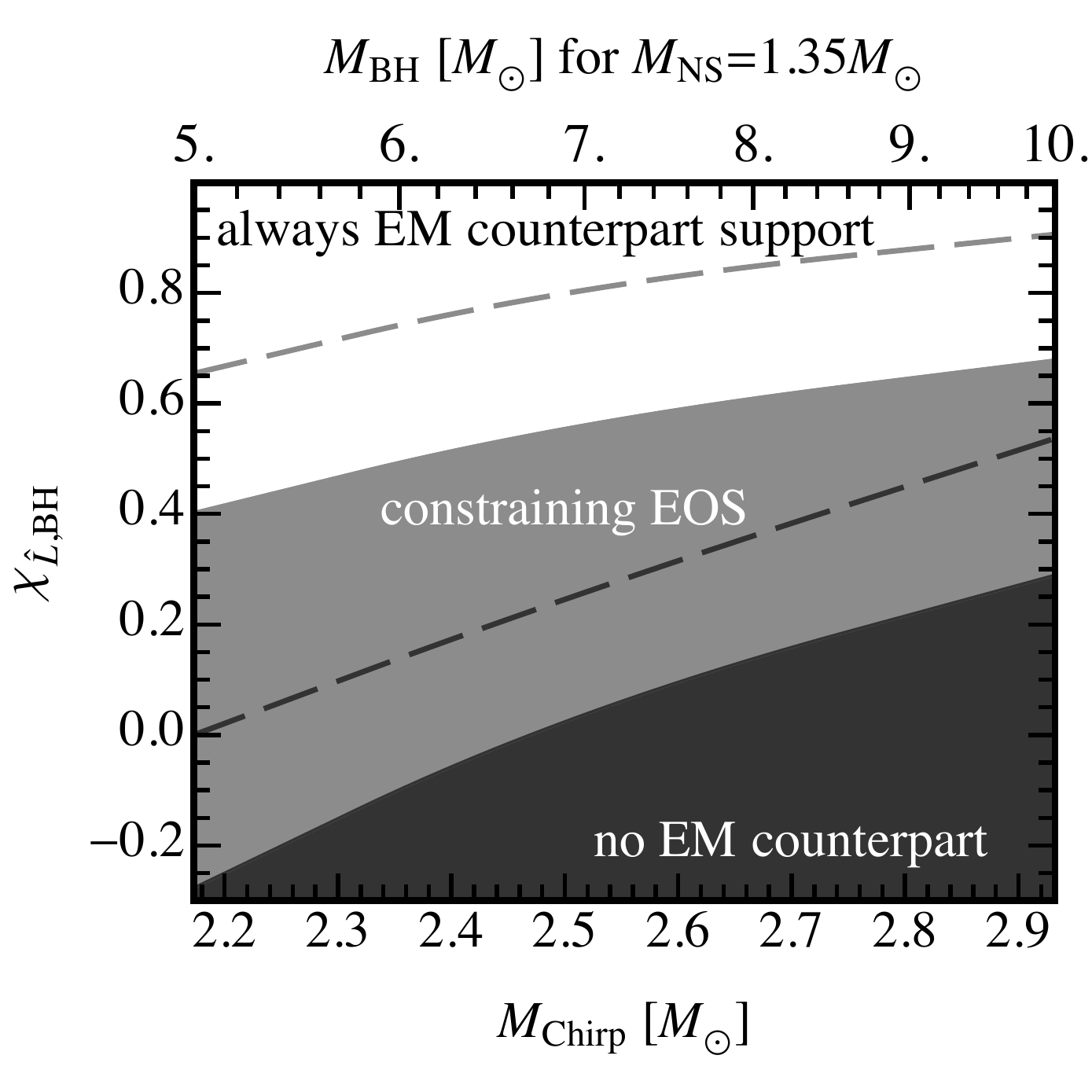}
  \caption{$\mChirp$--$\zspin$ combinations for which an \ac{EM}
    counterpart is not supported (dark gray), \ac{EM} counterpart
    support is \ac{EOS}-dependent (light gray), and no \ac{EOS}
    supports an \ac{EM} counterpart (white) for $\spin=0.998$.  Dashed
    lines denote the boundaries between the regions for aligned-spin
    binaries. \label{fig:summary}}
\end{figure}

Figure \ref{fig:summary} illustrates the parameter-space regions where
the above mentioned applications are potentially useful. In the
$\mChirp$--$\zspin$ plane --- the relevant physical quantities picked
up by the \ac{GW} \ac{PCA} --- we indicate for $\spin=0.998$ where
\ac{EM} counterparts are supported (white), where they are unlikely
(dark gray), and where their presence allows for constraining the
\ac{EOS}. For reference, we also include the region boundaries for
$\spin\equiv\zspin$ as dashed lines. Given an $\mChirp$, the
individual masses determine the $\zspin$ required to be in one of
those regions (Figure \ref{fig:realisticEOSs2}): we plot the minimal
$\zspin$ that allows for constraining the \ac{EOS} (lower contour) and
the minimal $\zspin$ that ensures that $\mDisk>\mThresh$ (upper
contour).

Throughout this Letter, we made very conservative assumptions about
our ability to measure parameters via \ac{GW} observations, and we
chose an almost extremal $\spin$ for a given $\zspin$, favoring the
onset of \acp{SGRB}. Consequently, we can conservatively estimate the
volume of parameter space where an \ac{EM} counterpart cannot be
ignited. Using a $2$H two-piecewise-polytrope \ac{EOS}
\citep{Kyutoku2010}, which gives a high maximum $\mNS$ of
$\sim2.8M_\odot$ and favors high $\mDisk$ values by yielding large
\ac{NS} radii, we span the intervals $\mNS\in[1,2.8]M_\odot$,
$\mBH\in[3,15]M_\odot$, and $\zspin\in [-0.95,0.95]$, keeping
$\spin=0.998$. We find that $\mDisk=0M_\odot$ in $\sim65$\% of the
volume of this parameter space, indicating that \emph{at most}
$\sim35$\% of the parameter space is useful for \ac{SGRB} trigger
follow-ups. This reduces to $\sim25$\% when considering aligned-spin
cases only, i.e., when $\spin\equiv\zspin$.

This has practical consequences for targeted \ac{GW} searches
following up \ac{EM} detections. Assuming an aligned-spin
template-based search, as advised by \citet{Canton:2014ena}, each
template covers a section of the parameter space, including precessing
binaries. We can now conservatively determine whether this section
could produce an \ac{EM} signature: combining our results with the
proper template-bank density \citep{Harry:2013tca}, we find that 43\%
of the templates in the above mentioned parameter space cover a region
with vanishing $\mDisk$. For non-precessing signals, this increases to
48\%. Hence, about half of the templates can be disregarded in a
search for \ac{SGRB} progenitors, promising an increase in both speed
and sensitivity.


\section*{Acknowledgments}
This work was supported by STFC grant No.~ST/L000342/1. We thank the
LIGO-Virgo collaboration CBC and GRB groups for their stimulating
environment, in particular Stephen Fairhurst, Raymond Frey, and Ian
Harry for interesting discussions throughout the genesis of this work
and for pushing the boundaries of this project further with their
questions. We also thank Patricia Schmidt and Mark Hannam for many
useful discussions about precessing binaries, and the anonymous
referee for helpful comments.

\bibliography{references}
\end{document}